# Study of iron complexes in visceral organs and brain from a $^{57}$Fe enriched β-thalassaemia mouse model via Mössbauer Spectroscopy

George Charitou[1)], Charalambos Tsertos[1)], Yiannis Parpottas[2)], Marina Kleanthous[3)], Carsten W Lederer[3)], Marios Phylactides[3)]

[1] Department of Physics, University of Cyprus, 1678 Nicosia, Cyprus

[2] School of Engineering, Frederick University, 1036 Nicosia, Cyprus

[3] Molecular Genetics Thalassaemia Department, The Cyprus Institute of Neurology and Genetics, 1683 Nicosia, Cyprus

**Abstract**

The hearts, kidneys, livers, spleens and brains of $^{57}$Fe enriched wild-type and heterozygous β-thalassaemic mice at 1, 3, 6 and 9 months of age were studied by means of Mössbauer Spectroscopy at 80K. Ferritin-like iron depositions in the heart and the brain of the thalassaemic mice were found to be slightly increased while significant amounts of Ferritin-like iron were found in the kidneys, liver and spleen. The Ferritin-like iron doublet, found in the organs, could be further separated into two sub-doublets representing the inner and surface structures of ferritin's mineral core. Surface iron sites were found to be predominant in the hearts and brains of all mice and in the kidneys of the wild-type animals. Ferritin rich in inner iron sites was predominant in the kidneys of the thalassaemic mice, as well as in the livers and in the spleens. The inner-to-surface iron sites ratio was elevated in all thalassaemic samples indicating that besides ferritin amount, the disease can also affect ferritin's mineral core structure.

## 1. Introduction

Iron is an essential bio-element, which is found in almost all forms of life because of its ability to act as an electron donor, Fe(II), and as an electron acceptor, Fe(III). In human metabolic process, iron serves as an oxygen carrier, as a transport medium for electrons within the cells, and as an integrated part of important enzyme systems in various tissues. In humans, iron mainly exists in complex forms, bound to protein as part of haem compounds (haemoglobin, myoglobin), haem enzymes, or non-haem compounds (flavin-iron enzymes, transferrin, and ferritin) [1].

Ferritin is the primary iron storage protein, which exists in both intracellular and extracellular compartments. It has a spherical protein shell (Apoferritin) with an 8-nm spherical cavity that can hold 2500-4500 atoms of iron(III) as a ferrihydrite mineral [2]. The shell is composed of 24 subunits of two types, the H and L-subunits, in various ratios that depend on the tissue type and can be modified by inflammatory and infectious conditions [2,3]. The H-subunit is found mostly in the heart, kidneys and brain [2,4] and its main role is the oxidation of Fe(II) iron to Fe(III), while the L-subunit assist in the core formation and is the predominant type in the liver and spleen [2]. For developing the mineral core in the human H-chain ferritin, there are the following three pathways [5]:

$$2Fe^{2+} + O_2 + 4H_2O \rightarrow 2Fe(O)OH_{(core)} + H_2O_2 + 4H^+ \tag{1}$$

$$2Fe^{2+} + H_2O_2 + 2H_2O \rightarrow 2Fe(O)OH_{(core)} + 4H^+ \tag{2}$$

$$4Fe^{2+} + O_2 + 6H_2O \rightarrow 4Fe(O)OH_{(core)} + 8H^+ \tag{3}$$

Once a sufficient core has been developed (≥800 Fe/Shell) from the Eqs. 1-2, Eq.3 becomes dominant and $Fe^{2+}$ is oxidized on the surface of ferritin's mineral core. Iron cores resulting from Eqs. 1-2 show similar MS characteristics [5].

Haemosiderin is also an iron storage complex which is found in conditions of iron overload [6] and is thought to be derived from the degradation of ferritin. It is water-insoluble with an amorphous structure and higher iron/protein ratio than ferritin, [6,7]. Haemosiderin also exists in normal conditions but in small amount.

Thalassaemias are hereditary anaemias resulting from defects in haemoglobin production. β-Thalassaemia, which is caused by a decrease in the production of β-globin chains, affects multiple organs and is associated with considerable morbidity and mortality. Accordingly, lifelong care is required. β-thalassaemia can be caused by any one of more than 200 mutations in the β-globin gene and is clinically heterogeneous because the different mutations affect β-globin chain synthesis to a different extent. The clinical severity of the disease depends on the number of mutated β-globin alleles and the severity of the mutation. The complications of thalassaemia include iron overload, bone deformities and cardiovascular illness. Currently, iron overload in transfused patients is responsible for most of the mortality and morbidity associated with thalassaemia. Iron deposition occurs in visceral organs (mainly in the heart, liver, and endocrine glands), causing tissue damage and ultimately organ dysfunction and failure. Cardiac events due to iron overload are still the primary cause of death in transfused individuals. Both transfusional iron overload and excess gastrointestinal absorption are contributory.

Mössbauer Spectroscopy (MS) is a versatile technique used to study the environment of a nucleus, with the absorption and re-emission of gamma rays. This technique uses a combination of the Mössbauer effect and Doppler shifts to probe the hyperfine transitions as a consequence of the interactions between the nucleus and its surrounding. It has been used to study many problems in physics, chemistry, biology and materials science. MS has been employed in earlier studies to investigate iron complexes in the blood and organs of humans and some animals. MS spectra derived from normal human liver and spleen samples [8] indicated the presence of iron in the form of ferrihydrite, together with some deoxyhemoglobin and methemoglobin in some samples. Tissue samples from iron overloaded livers [9] were also studied. In these livers, in addition to ferritin and hemosiderin, a new iron molecular environment typical of high spin ferric iron, characterized by a superparamagnetic behaviour, which begins at high temperature (above 77 K) was found. Ferritin and haemosiderin isolated from iron-overloaded human spleens has also been investigated by MS, at the temperatures of 1.3 and 200K [10]. The brain, liver and hearts of [57]Fe enriched healthy mice were studied by various researchers [11–13]. Haemoglobin from residual blood, ferritin, mitochondrial iron, and mononuclear non-haem high-spin (NHHS) Fe(II) and Fe(III) species were identified in brain samples [12]. The same components were also used to describe the livers of the mice in a later study [11]. Hearts from [57]Fe-enriched mice were also evaluated by MS [13]. Spectra consisted of a sextet and two quadrupole doublets. One doublet was due to residual blood, whereas the other was due to $[Fe_4S_4]^{2+}$ clusters and low-spin Fe(II) haem, most of which were associated with mitochondrial respiration. The sextet was due to ferritin. H-chain human ferritin with 500Fe/protein shell was also studied [5]. The spectra at 120K were fitted with the superposition of two doublets in nearly equal amounts, with similar isomer shifts but different quadrupole splittings. The two doublets where ascribed to iron spins at the surface and in the interior of the ferrihydrite nanoparticle.

However, no other previous MS work has utilized a thalassaemic animal model in order to study the major visceral organs and brain, and to investigate how the disease affects the formation and accumulation of iron complexes over time in the thalassaemic organs.

In a previous work [14], we demonstrated that the thalassaemic mouse model is a good candidate to study thalassaemia by means of Mössbauer spectroscopy. Subsequently, heterozygous thalassaemic (th3/+) and wild-type (C57BL/6) mice were enriched with $Fe^{57}$ [15] via gastrointestinal absorption to increase the statistics of the spectra.

In this follow-up paper, the results from the MS spectra of livers, spleens, kidneys, brains and hearts of $^{57}$Fe-enriched th3/+ and wild type (C57BL/6) mice at 1, 3, 6 and 9 months of age, as well as the brains of mice at 9 months of age, are presented.

## 2. Materials and methods

### 2.1. Experimental Setup

The experimental setup and the calibration process are described in Charitou et al. [14]. All parameter values extracted, are given relative to metallic iron at room temperature. The spectra were acquired at 80±0.5K. The max velocity was set at ~4.10mm/s with a second measurement at ~10mm/s, if it was deemed necessary.

### 2.2. Sample collection and preparation

For the purposes of this study, wild-type C57BL/6, and heterozygous thalassaemic (th3/+) mice were raised and enriched with $^{57}$Fe as described in Charitou et al. [15]. Due to the fact that MS measurements of bio-samples are time-consuming and also that the $^{57}$Fe isotope used for enrichment is fairly expensive, only one sample per group (th/3+ and wild-type) and age (1, 3, 6, 9 months) could be prepared and studied.

At the appropriate age, the mice were anaesthetized by intraperitoneal injection with tribromoethanol (10 μL/g) and the blood was extracted from the orbital sinuses. Exsanguination is crucial as peaks resulting from blood residues within the organs can hide weaker peaks arising from other iron complexes. The animals were then euthanized, and the heart, liver, kidneys, brain and spleen were carefully removed. The extracted organs were washed multiple times with Phosphate-Buffered Saline (PBS) and wiped thoroughly using clean sterile gauzes. Subsequently, they were weighted, placed in the appropriate polypropylene holders [14] and snap-frozen by submerging them into liquid nitrogen. The samples were stored into a liquid nitrogen dewar until needed for the measurement.

As the various organs have different shapes and sizes, holders of different diameters were used in order to eliminate empty space around the samples. For a specific type of organ, holders with same diameter were used in order to maintain comparability in terms of iron concentration (sub-spectrum's area of the fit over sample mass); the livers were placed in 12mm-diameter holders, the kidneys and brain in a 10mm-diameter holders and the hearts and spleens in 6mm-diameter holders. Some organs, like the heart and kidneys of mice at one month of age and the spleens from the wild-type mice of all ages, were very small and therefore the organs from two mice were combined in order to increase the mass of the sample. The th3/+ mice, similar to the human thalassaemia patients, suffer from splenomegaly, therefore their spleen was significantly enlarged and therefore only about half was placed into a holder. Table 1 presents the studied samples, the organ type, age of the mouse, holder diameter and sample mass.

**Table 1.** Sample details: mouse age, organ type, holder's diameter and sample mass

| Sample number | Group | Age (months) | Organ | Holder Diameter (mm) | Weight (g) |
|---|---|---|---|---|---|
| 1 | Wild-type | 1 | Heart | 6 | 0.21[1] |
| 2 | Wild-type | 1 | Kidneys | 10 | 0.60[1] |
| 3 | Wild-type | 1 | Liver | 12 | 0.84 |
| 4 | Wild-type | 1 | Spleen | 6 | 0.11[1] |
| 5 | Th3/+ | 1 | Heart | 6 | 0.19[1] |
| 6 | Th3/+ | 1 | Kidneys | 10 | 0.45[1] |
| 7 | Th3/+ | 1 | Liver | 12 | 0.96 |
| 8 | Th3/+ | 1 | Spleen | 6 | 0.19 |
| 9 | Wild-type | 3 | Heart | 6 | 0.23[1] |
| 10 | Wild-type | 3 | Kidneys | 10 | 0.50 |
| 11 | Wild-type | 3 | Liver | 12 | 1.04 |
| 12 | Wild-type | 3 | Spleen | 6 | 0.12[1] |
| 13 | Th3/+ | 3 | Heart | 6 | 0.25 |
| 14 | Th3/+ | 3 | Kidneys | 10 | 0.38 |
| 15 | Th3/+ | 3 | Liver | 12 | 1.24 |
| 16 | Th3/+ | 3 | Spleen | 6 | 0.20 |
| 17 | Wild-type | 6 | Heart | 6 | 0.21 |
| 18 | Wild-type | 6 | Kidneys | 10 | 0.17 |
| 19 | Wild-type | 6 | Liver | 12 | 0.81 |
| 20 | Wild-type | 6 | Spleen | 6 | 0.17[1] |
| 21 | Th3/+ | 6 | Heart | 6 | 0.27 |
| 22 | Th3/+ | 6 | Kidneys | 10 | 0.39 |
| 23 | Th3/+ | 6 | Liver | 12 | 0.93 |
| 24 | Th3/+ | 6 | Spleen | 6 | 0.27 |
| 25 | Wild-type | 9 | Heart | 6 | 0.21 |
| 26 | Wild-type | 9 | Kidneys | 10 | 0.55 |
| 27 | Wild-type | 9 | Liver | 12 | 1.35 |
| 28 | Wild-type | 9 | Spleen | 6 | 0.21[1] |
| 29 | Wild-type | 9 | Brain | 10 | 0.52 |

| 30 | Th3/+ | 9 | Heart | 6 | 0.18 |
| 31 | Th3/+ | 9 | Kidneys | 10 | 0.32 |
| 32 | Th3/+ | 9 | Liver | 12 | 0.75 |
| 33 | Th3/+ | 9 | Spleen | 6 | 0.20 |
| 34 | Th3/+ | 9 | Brain | 10 | 0.42 |

[1]Organs from two mice were used

### 2.3. Spectra fitting methodology

Initially, all spectra were fitted with free Lorentzian peaks. From this, it was observed that four similar doublets could be used to fit most of the spectra, while in two spectra a sextet could also be observed. Due to the strong similarities among the sub-doublets needed to fit the data, the mean values for these components were calculated for each group of mice. As no significant differences were observed in the MS parameters between the thalassaemic and wild-type mice, the overall mean values were calculated and subsequently all spectra were re-fitted with the values seen in Table 2. For consistency, the quadrupole splitting $(\Delta Eq)$ was kept fixed, while the width $(\Gamma)$, and isomer shift $(\delta)$, was left to vary slightly, if needed.

**Table 2.** The mean values used for the final fit of the spectra. They were calculated based on initial fits with free parameters (N: the number of the different spectra used to calculate the mean value, wt: wild-type mice, th3/+: thalassaemic mice)

| Iron Complex | | N | $\delta$ (mm/s) | $\Delta Eq$ (mm/s) | $\Gamma$ (mm/s) | HMF (T) |
|---|---|---|---|---|---|---|
| High spin Fe(II) haem (Deoxy-Hb) | Average for wt | 17 | 0.91 | 2.28 | 0.32 | |
| | Average for th3/+ | 17 | 0.91 | 2.28 | 0.32 | |
| | Average | 34 | 0.91 | 2.28 | 0.32 | |
| NHHS Fe(II) | Average for wt | 6 | 1.35 | 2.83 | 0.45 | |
| | Average for th3/+ | 9 | 1.36 | 2.94 | 0.41 | |
| | Average | 15 | 1.36 | 2.90 | 0.43 | |
| Ferritin-like (I) | Average for wt | 13 | 0.45 | 0.59 | 0.40 | |
| | Average for th3/+ | 17 | 0.46 | 0.59 | 0.42 | |
| | Average | 30 | 0.46 | 0.59 | 0.42 | |
| Ferritin-like (II) | Average for wt | 17 | 0.46 | 1.06 | 0.41 | |
| | Average for th3/+ | 17 | 0.45 | 1.05 | 0.45 | |
| | Average | 34 | 0.46 | 1.05 | 0.43 | |
| Haemosiderin | Average th3/+ | 2 | 0.48 | -0.25 | 1.04 | 45.91 |

## 3. Results and Discussion

### 3.1. High Spin Fe(II) haem iron

High Spin Fe(II) haem iron with fitting parameters δ=0.91 mm/s and ΔEq=2.28 mm/s was found in all samples studied. This iron exhibits similar parameters with deoxy-haemoglobin and its presence is probably due to trapped blood residues that could not be removed during the exsanguination and rinsing with Phosphate-Buffered Saline (PBS).

### 3.2. NHHS Fe(II) complexes

The second component found in some organs showed broadened linewidth and it was fitted with δ=1.36 mm/s and ΔEq=2.90 mm/s parameters. Due to the fact that this MS doublet is weak (because of its low concentration), and it has the peak of lowest energy well hidden in more intense components, the exact MS values must be taken with some scepticism. A similar component was found in mice livers [11] and brains [12] and it was attributed to Mononuclear Non-haem high-spin Iron (NHHS Fe(II)). NHHS Fe(II) complexes can have labile ligands that undergo Fenton chemistry, which are responsible for the creation of reactive oxygen species (ROS) [11,12] and are known to cause extensive cellular damage. In this study, NHHS Fe(II) was found in the kidneys, livers and brain of the healthy and th3/+ mice of all ages, and in the spleen of the adult (3, 6, 9 months of age ) th3/+ mice only. Figure1 shows the average NHHS Fe(II) concentration (doublet's area per sample mass) for the brain, kidneys, liver and spleen samples of the th3/+ and wild-type adult mice (3, 6, 9 months of age). The liver and kidneys of the adult th3/+ mice had on average 14% and 500% more NHHS Fe(II) per sample mass, respectively, compared to the corresponding healthy mice. In the spleens of the th3/+ adult mice, NHHS Fe(II) iron was found in high concentration, while there was no sign of NHHS Fe(II) iron in spleen samples from the control group.

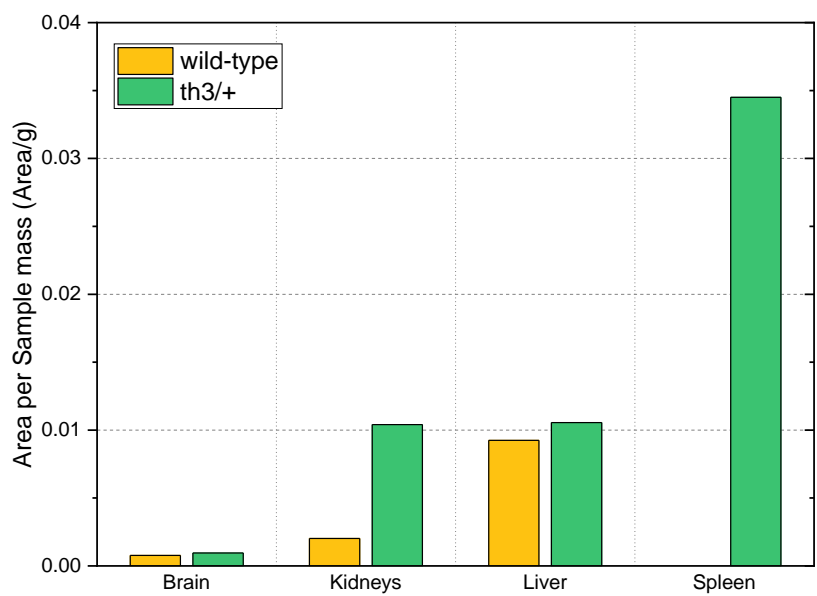

**Fig. 1.** NHHS Fe(II) average concentration (doublet's area per sample mass) for the brain, kidneys, liver and spleen samples of th3/+ and wild-type adult mice (3, 6, 9 months of age). Note that the concentration might not be comparable between the samples of different organs due to differences in the diameter of the prepared absorbers.

### 3.3. Ferritin-like iron

A major feature in all spectra, especially the thalassaemic ones, was a wide doublet with δ≈0.46 mm/s and ΔEq≈0.74 mm/s. This doublet was observed in previous Mössbauer studies, in samples of various organs from humans and animals, and it was attributed to ferritin-like iron [8–10,16]. The use of one doublet to fit this ferritin-like iron (as in most previous studies) was inadequate, therefore two sub-doublets were used which improved the fitting. These two sub-doublets showed fitting parameters of δ=0.46 mm/s and ΔEq=0.59 mm/s, and δ=0.46 mm/s and ΔEq= 1.05 mm/s, respectively, as seen in Table 2. These parameters are similar to the ferritin doublets found by Bou-Abdallah et al. [5].

Bou-Abdallah et al. [5] after studying the cores of human H-chain ferritin with 500 Fe/shell, described their spectra by the superposition of two magnetic substances in nearly equal amounts that arise from the ferritin's mineral core. At 120K, the two substances are presented by two doublets which give rise to magnetic ones at lower temperatures. The parameters of the doublets between 4.2-120K are δ=0.46 mm/s (for both substances), and ΔEq of 0.55-0.59 mm/s and 1.01-1.04mm/s for each one, respectively. According to the authors, the doublet with the smaller ΔEq corresponds to the iron sites at the interior of the mineral core while the second one to the iron sites at the surface of the core. Even though our MS doublet parameters are similar to the ones found in Bou-Abdallah et al. [5], the ratios of our sub-doublets are not equal to 1 as in Bou-Abdallah et al. [5]. In particular, it was found that this ratio depends on the organs and the type of mouse. In the wild-type mice, the hearts, kidneys and brain showed higher concentration of surface iron sites, while the spleen and liver showed higher concentration of inner iron sites. This pattern is similar with the one expected from the light (L) and heavy (H) chains of ferritin in the organs. L-subunits are predominant in tissues with high levels of stored iron, like spleen and liver, and H-subunits are predominant in tissues with no iron storage function, like the heart [17]. Based on this, the first sub-doublet (inner iron sites) might relate to the L-chains and long-term iron storage, while the second sub-doublet (surface iron sites) to the H-chains and the ferroxidase activity of ferritin.

Besides the work by Bou-Abdallah et al. [5], two other studies by Holmes-Hampton et al. [12] and Chakrabarti et al. [11] also used two doublets to describe this area of the spectrum from the liver and brain of healthy mice, but with different parameters. The authors attributed their sub-doublets to ferritin and to mitochondrial respiratory complexes ([$Fe_4S_4$]$^{2+}$ clusters and low-spin Fe(II) haem centres). Fitting our spectra with similar sub-doublets as in these studies was attempted but did not provide a satisfying fit.

### 3.3.1. Ferritin-like iron in murine cardiac tissue

MS spectra from the hearts of wild-type and th3/+ mice at 9 months of age are shown in Figs. 2a and 2b, respectively. In the cardiac tissues, the surface iron sites were found to be the predominant type for the wild-type and th3/+ groups of mice of all ages. Their concentration (doublet's area per sample mass) of the surface iron sites seems to increase only slightly with age while the concentration of the inner iron sites increases by a greater degree, as seen in Fig. 3a and 3b. Because of this, the inner-to-surface iron sites (I/S) ratio increases in favour of the inner iron sites and with a similar rate for both groups (Fig 4). Based on the measurements, juvenile th3/+ mice (at 1 month of age) begin their lives with an increased total ferritin-iron levels (sum of the area of both iron sites) in the cardiac tissue compared to the wild-type ones, but the latter seem to be raised at equal levels with the th3/+ ones at the age of 9 months (Fig. 3a and 3b). Thalassaemia patients without iron chelation treatment present significantly higher iron amounts in the cardiac tissues than the healthy individuals. This is in contrast with the above mentioned observation. A difference in the pattern of iron accumulation in the heart between thalassaemic mice and humans was also observed by Gardenghi et al. [18]. Gardenghi et al.

[18] suggested that the pattern of iron accumulation in the heart may differ between mice and humans or the underlying disease in mice may enhance the effect of the small increases in heart iron content.

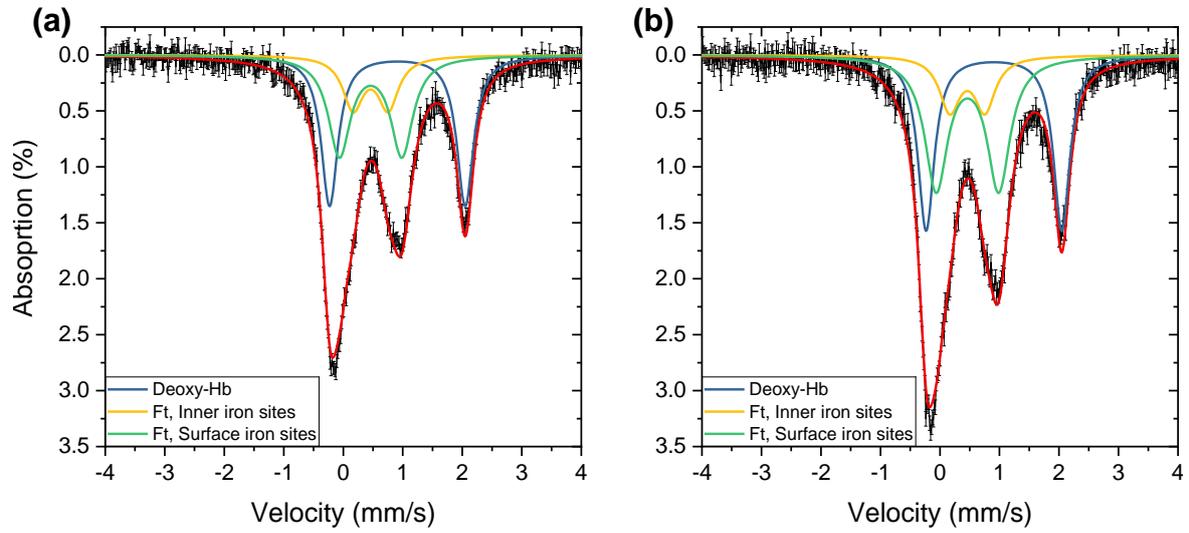

**Fig. 2.** MS spectra from the hearts of mice at nine months of age. The spectrum from the wild-type mouse is shown in part **(a)** and from the th3/+ in part **(b),** respectively.

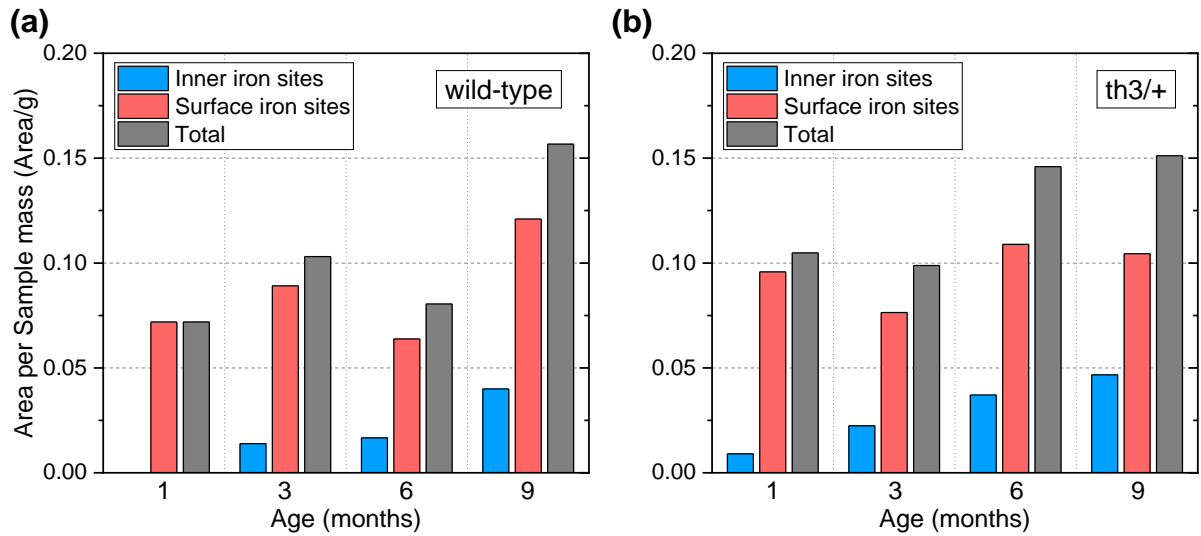

**Fig. 3.** Ferritin-like iron concentration in the cardiac tissue of samples from wild-type **(a)** and th3/+ **(b)** mice

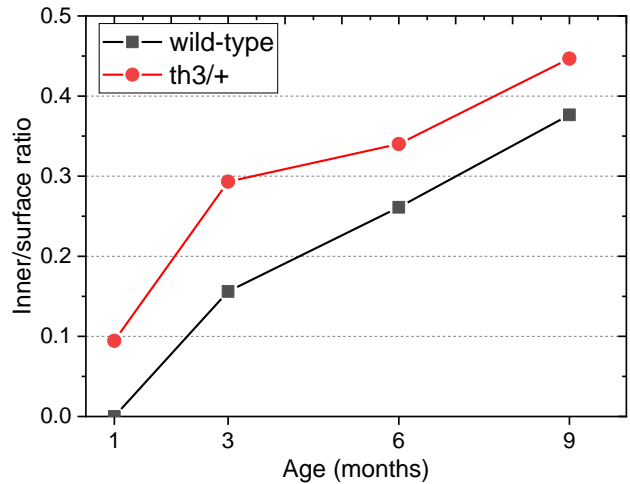

**Fig. 4.** The trend for the inner-to-surface iron sites ratio as calculated from the heart samples of wild-type and thalassaemic th3/+ mice of different ages. The statistical errors are < 5%.

### 3.3.2. Ferritin-like iron in murine renal tissue

In contrast to the cardiac tissues, where there was little difference in terms of iron accumulation between wild-type and th3/+ mice, the effect of thalassaemia on iron accumulation in the kidneys of the th3/+ mice was more severe, especially for the adult mice. This resulted to significantly increased absorption in the MS spectra, as shown in the two spectra from mice at 9 months of age (Fig 5a and 5b).

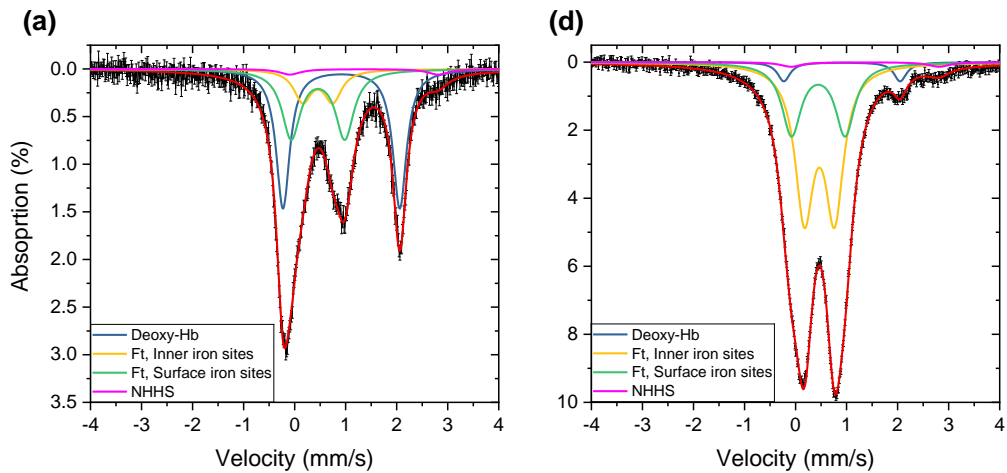

**Fig. 5.** Spectra from the kidneys of mice at 9 months of age. The spectrum from the wild-type mouse is shown in part (a) and from the th3/+ in part (b), respectively. Note the difference in the absorption scale.

In the renal tissues of all wild-type mice, the predominant ferritin iron sites are the surface-type ones (Fig 6), with a similar inner-to-surface ratio as in the cardiac tissues. The juvenile thalassaemic th3/+ mouse at 1 month of age still had predominantly surface-type iron sites, but the I/S ratio was higher (I/S ratio = 0.55), when compared to the wild-type animals (I/S ratio = 0.13-0.42). At 3 months of age, the concentration of the inner iron sites increases dramatically and becomes the predominant type of iron-sites with an IS ratio above 1 (I/S ratio = 1.5). Furthermore, the I/S ratio continued to increase with the thalassaemic mice age (Fig. 6). These observations suggest that thalassaemic mice tend to favour the accumulation of ferritin rich inner iron sites, in their renal tissues.

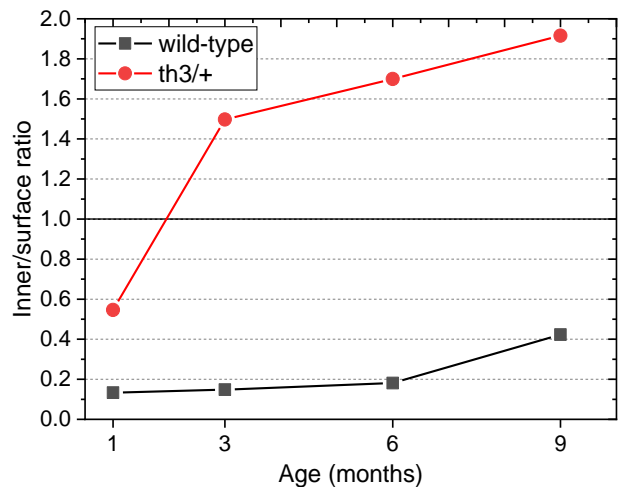

**Fig. 6.** The trend for the inner-to-surface ferritin's iron sites ratio calculated from the kidney samples of wild-type and thalassaemic th3/+ mice. The statistical errors are < 5%.

Relative to the total iron concentration in the kidneys, the thalassaemic mouse at 1 month of age showed 2.4 times higher values than the wild-type equivalent. In the adult mice, renal iron accumulation is much more severe, as seen in Fig. 7a and 7b where the iron concentration in the renal tissues of wild-type and th3/+ mice is presented, with the th3/+ mouse at 6 months of age having 15.7 times higher iron concentration than the wild-type one. On an average, the adult th3/+ mice showed ~11.6 times higher iron concentration than the control group.

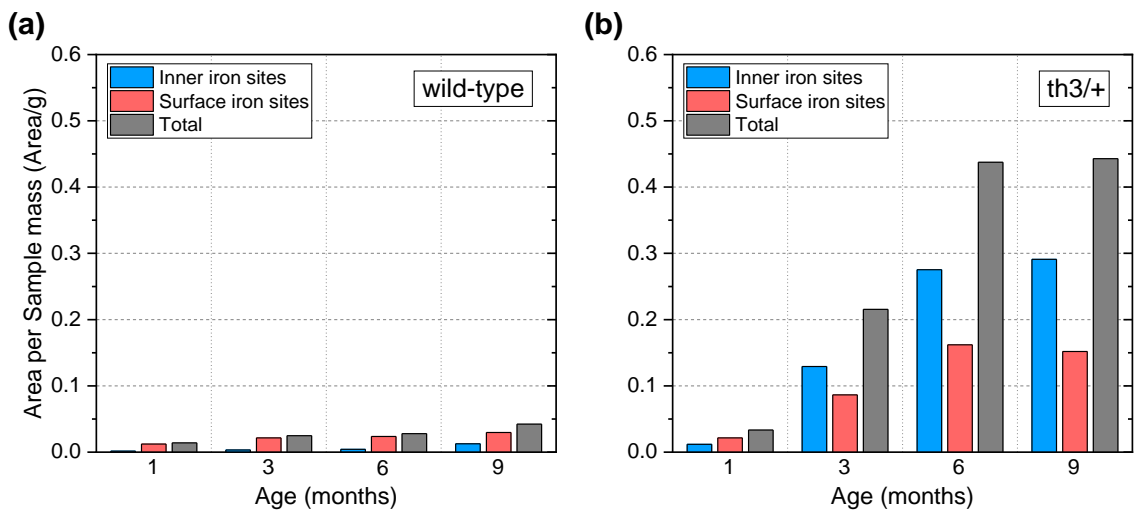

**Fig. 7.** Ferritin-like iron in the kidney samples of wild-type **(a)** and thalassaemic th3/+ **(b)** mice. A huge increase in both ferritin iron sites can be seen in the th3/+ samples

### 3.3.3. Ferritin-like iron in murine liver tissue

In the juvenile wild-type mice, only surface iron sites were found while in the adult mice the inner iron sites were the predominant ones. This can be seen in Fig. 8a and 8c. The overall ferritin levels of the wild-type adult mice (3-9 months of age) do not seem to change significantly with age (Fig. 8a) nor the inner-to-surface ratio. The juvenile thalassaemic th3/+ mouse at 1 month of age exhibited equal amounts of the two types of iron sites, and had total ferritin levels that were 2.4 times higher than the control equivalent. In the adult th3/+ mice, the predominant ferritin iron sites were again the inner-type, with slightly increased inner-to-surface ratio (Fig 8b and 8c). In contrast to the control samples, the total ferritin concentration in the thalassaemic samples was found to increase linearly over time (Fig. 8d) with the samples from the adult th3/+ mice having 4 - 9.5 times more ferritin iron than the same-age controls. MS spectra from the liver samples of mice at 9 months of age are shown in Fig. 9a and 9b.

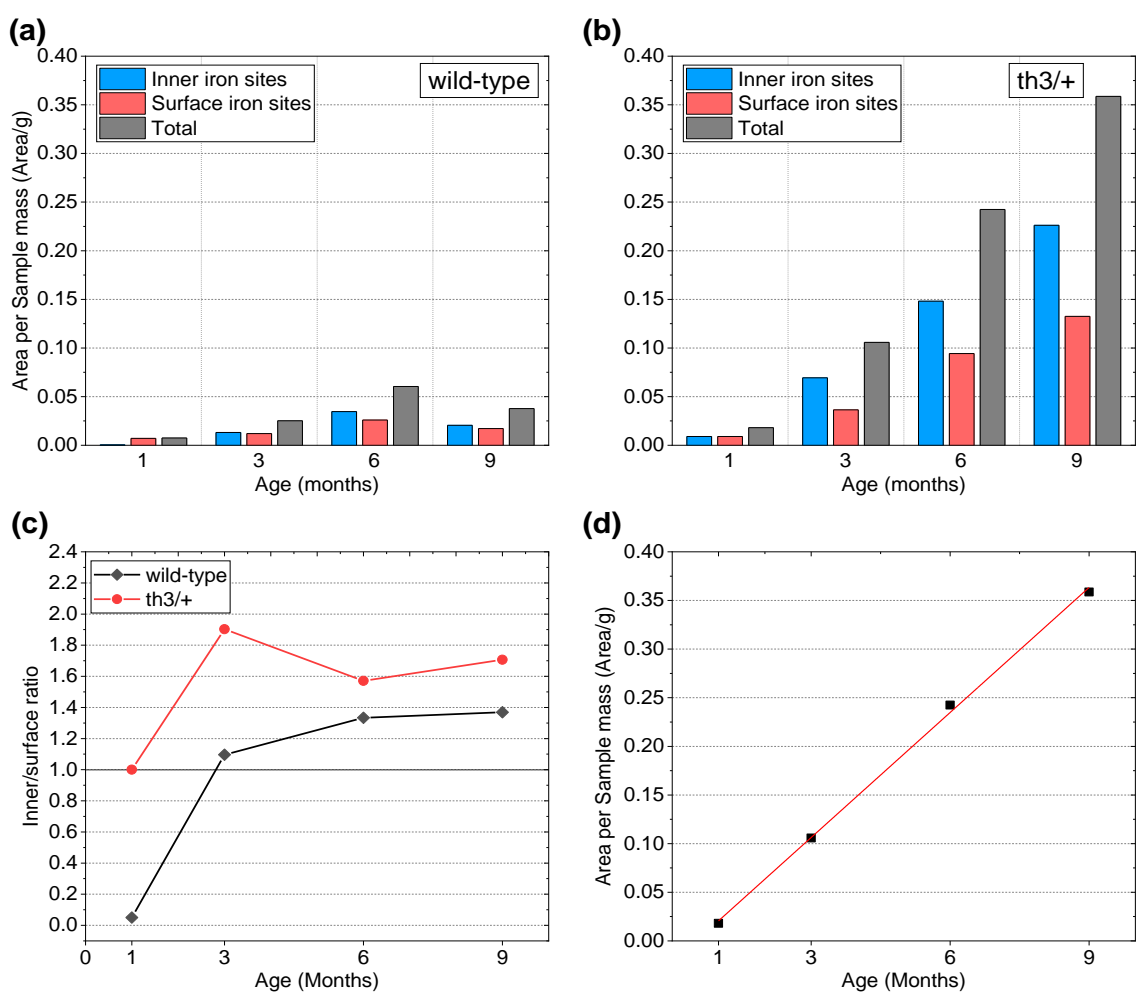

**Fig. 8.** Ferritin-like iron in the liver samples of wild-type **(a)** and thalassaemic th3/+ **(b)** mice. The trends for the inner-to-surface ferritin's iron sites ratios in the liver **(c)** and linear correlation of liver's total ferritin-like iron with the mouse age in the th3/+ **(d)**. The statistical errors in (c) and (d) are $< 5\%$.

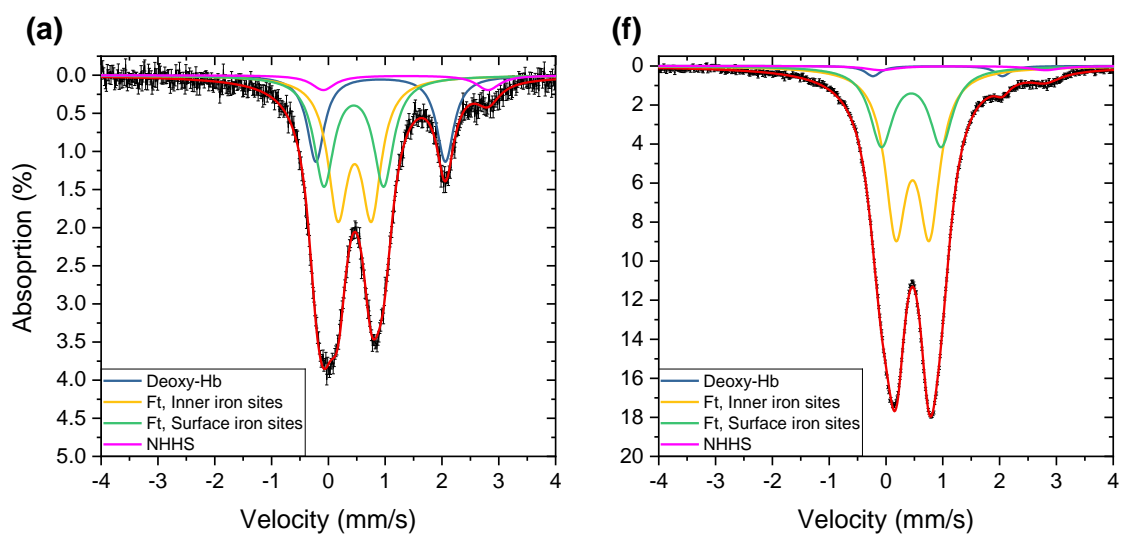

**Fig. 9.** MS spectra from the livers of mice at 9 months of age: wild-type **(a)** and th3/+ **(b).** Note the difference in the absorption scale

### 3.3.4. Ferritin-like iron in murine splenic tissue

The spectra from the spleen samples of th3/+ mice at 6 and 9 months of age show an extra component not observed in the previous samples. The samples were measured with MS for a second time and spectra were taken at higher velocity ($V_{max} = \pm 10$ mm/s). This extra component is seen as a wide sextet (Fig. 10b), with fitting parameters of δ= 0.48 mm/s, ΔEq= -0.25 mm/s, HMF= 45.91 T and it can be attributed to haemosiderin with mineral goethite structure [19]. Haemosiderin is an iron storage complex found mainly in iron-overload conditions, and was observed in iron overloaded spleen tissues of thalassaemia patients [10,19] and iron overloaded rats [20]. From this observation, we can fairly safely assume that the th3/+ mice at 6 and 9 months of age exhibited advanced iron-overload conditions. Spectra from the spleen samples of mice at 9 months of age are presented in Figs. 10a and 10b, respectively.

Similar to the liver tissues, the predominant type of ferritin iron sites in the spleen samples was found to be the inner-type for both groups and all ages of mice. All adult mice show similar inner-to-surface ratio (Fig. 11c) with a mean value of 2.34:1 (~70% inner-core ferritin). The total iron (ferritin and haemosiderin) concentration increases with age for both groups of mice as seen in Fig. 11a and 11b. However, th3/+ mice seem to deposit iron in their spleen in a higher rate than the wild-type ones. Overall, the thalassaemic samples showed about 2.5~3 times more ferritin and haemosiderin iron per sample mass than the healthy controls.

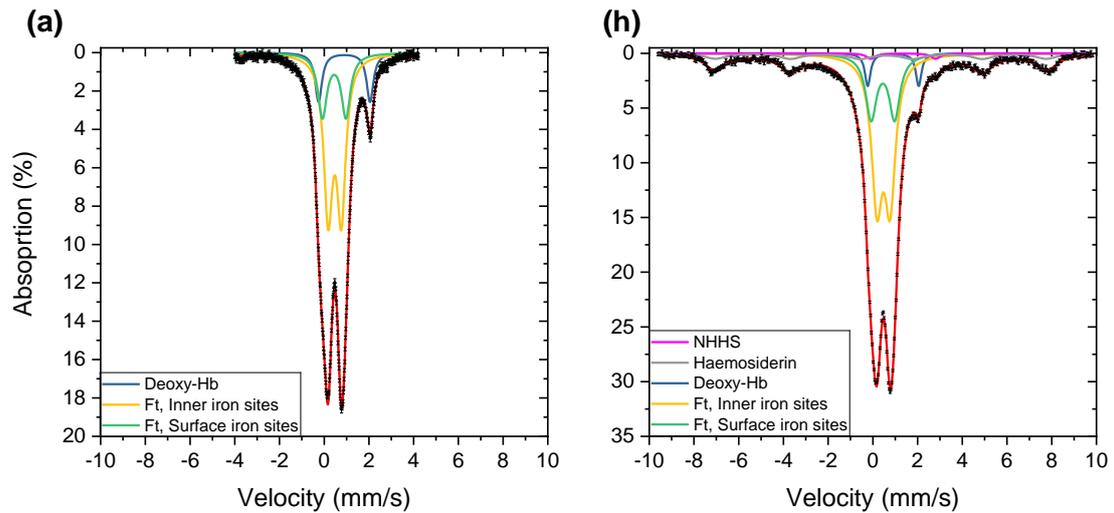

**Fig. 10.** Mössbauer spectra from the spleens of wild-type **(a)** and th3/+ mice **(b)** at 9 months of age. The hemosiderin sextet, a sign of iron overload, is clearly seen in the th3/+ sample. The spectrum from the thalassaemic mouse was acquired at higher velocity due to the sextet

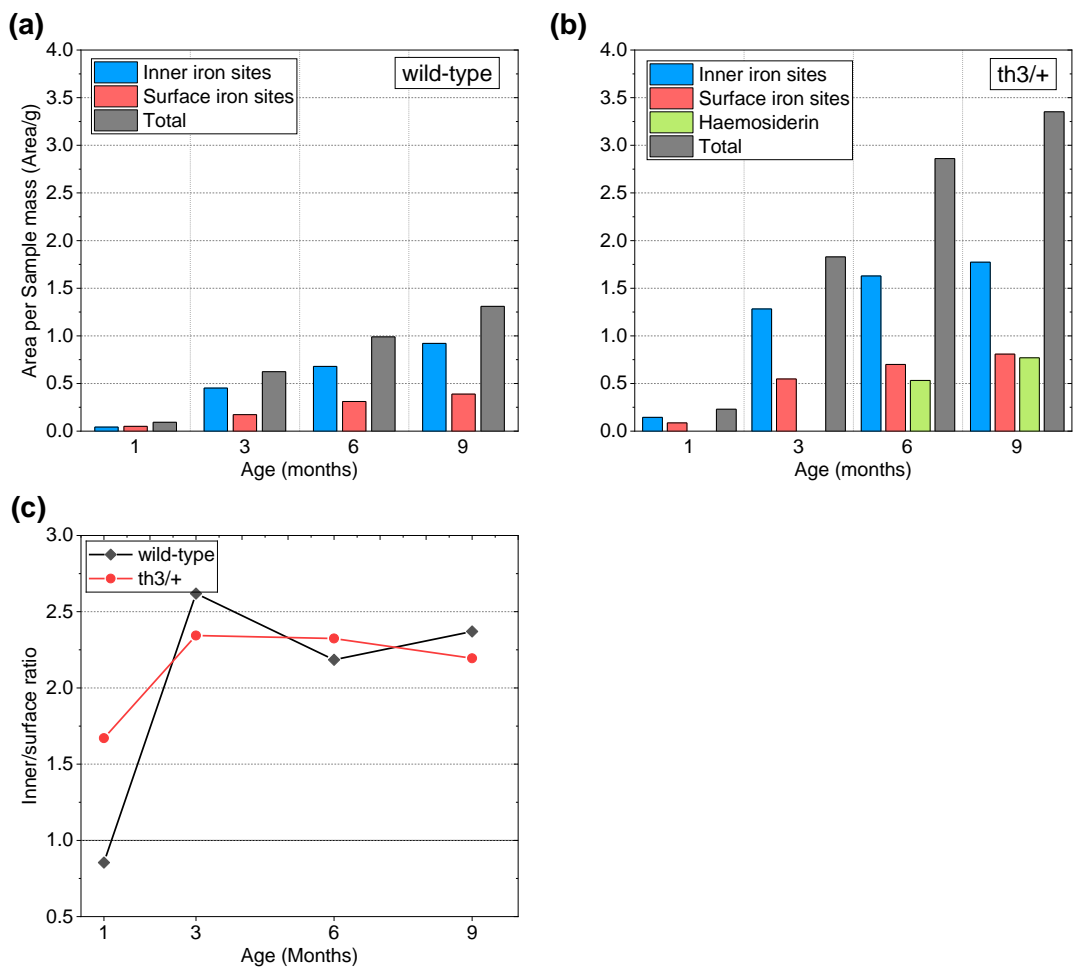

**Fig. 11.** Ferritin-like iron in the spleen samples of wild-type **(a)** and thalassaemic th3/+ **(b)** mice. In the th3/+ mice of 6 and 9 months of age, a new iron complex (green) can be seen which can be attributed to haemosiderin. The trend for the inner-to-surface ferritin iron sites ratio in the spleen samples **(c)**. The ratio for the adult mice seems to be independent of the sample group. The statistical errors in (c) are < 5%.

### 3.3.5. Ferritin-like iron in murine brain tissue

In order to investigate if there are indications of increased iron accumulation in the brain as a result of thalassaemia, brain tissues from mice at 9 months of age were studied. It should be noted that some neurodegenerative diseases such as Alzheimer's and Parkinson's might be correlated to iron presence in the brain [21,22]. Therefore, as thalassaemia patients suffer from an increased iron deposition in various organs, it is interesting to also study the brains. Increased iron levels or changes in the nature of iron complexes might indicate new medical problems that thalassaemia patients might face as they age. The measured MS spectra are shown in Fig. 12a and 12b, respectively. Note that, even though the absorption in these spectra is relatively small (<1%), the overall measured background, mainly due to iron presence in the detector's window, was found to be <0.1% and therefore, it is not expected to affect the results significantly.

In the samples under investigation, the predominant ferritin iron sites in both groups of mice was found to be the surface-type, as seen in Fig. 13. Also similar to the cardiac and renal tissues, the th3/+ mouse showed increased inner iron sites and thus elevated inner-to-surface ratio (I/S=0.81) compared to the wild-type ones (I/S = 0.61). Even though, there is an increased absorption in the spectrum from the brain of the wild-type mouse due to deoxyhaemoglobin, compared to the thalassaemic one, the thalassaemic sample presented ~10% more ferritin iron than the wild-type equivalent. However, no conclusive statements can be further made concerning the amount of ferritin in the brain, due to the small difference in ferritin percentage found, and the small sample population.

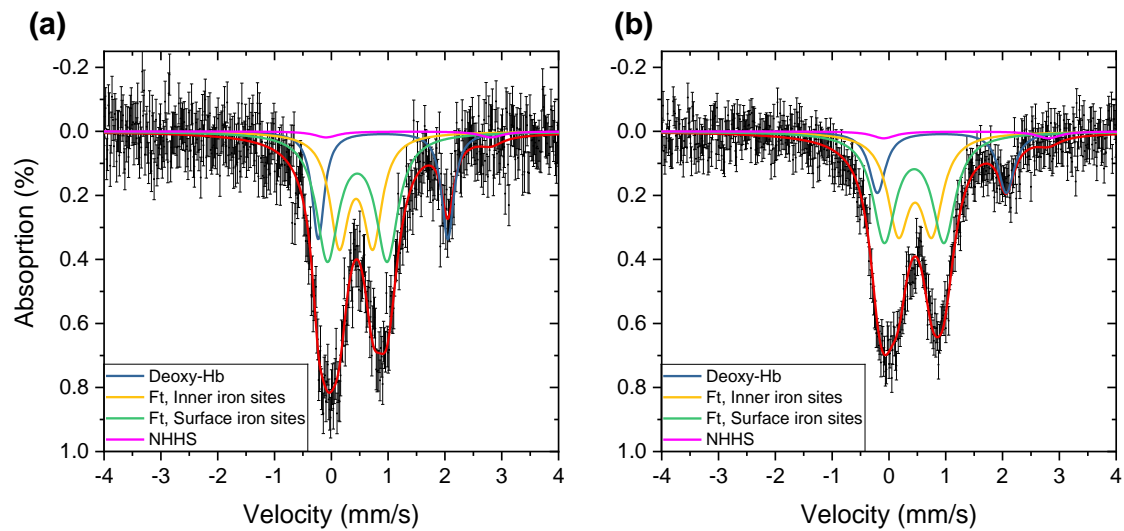

**Fig. 12.** Mössbauer spectra from the brains of a wild-type **(a)** and a th3/+ mice **(b)** at 9 months of age

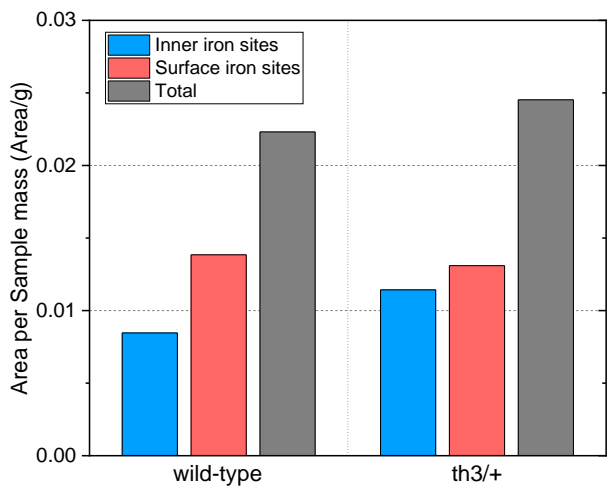

**Fig. 13.** Ferritin-like iron in the brain samples of one wild-type and one th3/+ mouse at 9 months of age

## 4. Conclusions

Mössbauer spectra from hearts, kidneys, livers, spleens and brains from wild-type C57BL/6 and thalassaemic th3/+ $^{57}$Fe enriched mice at 1, 3, 6, and 9 months of age, were obtained at 80K in order to study the iron complexes and iron accumulation in the major visceral organs and brain as a result of thalassaemia.

Overall, iron complexes of four iron "categories" were identified. A doublet from high-spin Fe(II) iron with deoxy-haemoglobin parameters was found in all samples and was attributed to blood residues inside the tissues studied. The second component was attributed to Mononuclear Non-haem high-spin Iron (NHHS Fe(II)). These complexes can undergo Fenton chemistry due to labile ligands, therefore they can be responsible for the creation of reactive oxygen species, which is a major issue for thalassaemia patients and responsible for cell and eventually organ damage. NHHS Fe(II) iron was found in elevated values in the brain, kidneys and the liver of the th3/+ mice compared to the wild-type animals. Also, high amounts of NHHS Fe(II) iron were found in the spleen of adult th3/+ mice only. Th3/+ mice suffer from splenomegaly, a condition that thalassaemia patients also suffer from, where the spleen enlarges greatly in size due to increased extramedullary erythrocyte production and destruction. Thus, it is possible that the increased NHHS Fe(II) concentration in the thalassaemic spleen tissue might be related to phenomena associated with erythropoiesis and/or erythrophagocytosis.

Besides the NHHS Fe(II) and deoxy-haemoglobin, ferritin-like iron was found in all organs, as expected. The wide ferritin-like doublet found in many previous studies could be further fitted with two sub-doublets which relate to the ferritin mineral core structure, and more specifically to the iron sites on the surface and in the centre of the core. The ratio of these two sub-doublets was found to depend greatly on the organ type and the mouse genotype. In the wild-type mice, (a) the hearts, kidneys and brains, organs with no long-term iron storage function, showed greater amounts of surface iron sites, while (b) the liver and the spleen contained predominantly inner-core iron sites. A correlation with the L and H chains of ferritin might be possible, as cardiac tissues are rich in H-chain ferritin, while liver and spleen are rich in L-chain ferritin. Thalassaemic th3/+ mice favoured the formation of inner iron sites. Based on these two observations, it can be postulated that when large amounts of iron are present, as in the case of thalassaemia, they are stored in the centre of the ferritin mineral core and not on its surface. In addition, the two chemical pathways responsible for the production of the inner-core, either produce (Eq. 1) or utilize (Eq. 2) hydrogen peroxide ($H_2O_2$), which is a major contributor to oxidative damage.

In the kidneys of the thalassaemic mice, significant amounts of ferritin-like iron depositions were found when compared to the wild-type equivalents. The th3/+ mice at 6 months of age showed ~16 times more ferritin-like iron in the renal tissues than the healthy controls. This finding indicates that kidneys might accumulate iron in great amounts and hence be vulnerable to iron overload leading to new complications for the thalassaemia patients that can face later on, in life.

In the two examined brains from mice at 9 months of age, around 10% more ferritin-like iron was found in the th3/+ mouse. This might be of great interest as brain iron might be linked to various neurodegenerative diseases. In the cardiac tissues, small increases in iron concentration were found, especially for the juvenile th3/+ mouse. However, the wild-type mouse at 9 months of age revealed iron concentration similar to the thalassaemic one. This observation is not consistent with the thalassaemic patients, as the iron concentration in their cardiac tissues increases significantly over time. Thalassaemic murine liver tissues investigated showed significant increases in iron concentration, up to ~10 times the normal amount, which was expected to some degree, as iron accumulation in the liver is a well-known complication in thalassaemia patients. The rate at which iron accumulates in the th3/+ livers was found to be correlated linearly with mouse age.

Haemosiderin, an iron storage complex that is found in the spleen of thalassaemic and iron overloaded patients, was also observed in the spleens of the thalassaemic th3/+ mice at 6 and 9 months of age. This indicates that these mice suffered from advanced iron-overload conditions even though the iron content in their diet was at relative low levels. This can be useful for the planning of future studies in order to decrease both the required time and the expenses.

In respect to iron accumulation with mice age, th3/+ thalassaemic mice continuously accumulate significant amounts of ferritin-like iron in their liver, kidneys and spleen up to the age of 9 months, which was the maximum age considered in this study. On the other hand, ferritin-like iron concentration in the cardiac tissue is increased in the juvenile thalassaemic th3/+ mice when compared to the wild-type ones. With age, and in contrast to the other organs, iron accumulation in the cardiac tissue of the thalassaemic mice seems to decrease, and at the age of 9 months, both mice groups exhibit similar ferritin-like iron concentration levels.

In conclusion, this study has shown that thalassaemic mice accumulate significant amounts of iron as they aged, mainly in the form of ferritin-like iron, in their liver, kidneys and spleen compared to the wild-type ones. This study has also shown that the structure of the ferritin's mineral core varies according to the implicated tissues. Therefore, valuable information about iron complexes and their concentration in these organs can be extracted by studying $^{57}$Fe enriched mice with Mössbauer Spectroscopy.

**Acknowledgments:** We gratefully acknowledge the personnel of the Transgenic Mouse Facility of the Cyprus Institute of Neurology and Genetics for the planning of the mouse breeding. This work was funded by a UCY grant (8037P-3/311/23030).

**Ethical Approval:** All applicable international, national and/or institutional guidelines for the care and use of animals were followed.